\documentclass[prd,twocolumn,a4paper,amsmath,amssymb]{revtex4} 
\usepackage{graphicx}

\begin{document}
\title{Hydropathy Conformational Letter and its Substitution Matrix \\
HP-CLESUM: an Application to Protein Structural Alignment }
\author{Sheng~Wang$^{1,*}$}
\affiliation{$^1$Institute of Theoretical Physics, Chinese Academy
of Sciences, Beijing 100190, China}

\begin{abstract}
\begin{footnotesize}
{\bf Motivation:} Protein sequence world is discrete as 20 amino
acids (AA) while its structure world is continuous, though can be
discretized into structural alphabets (SA). In order to reveal the
relationship between sequence and structure, it is interesting to
consider both AA and SA in a joint space. However, such space has
too many parameters, so the reduction of AA is necessary to bring
down the parameter numbers.

{\bf Result:} We've developed a simple but effective approach called
entropic clustering based on selecting the best mutual information
between a given reduction of AAs and SAs. The optimized reduction of
AA into two groups leads to hydrophobic and hydrophilic. Combined
with our SA, namely conformational letter (CL) of 17 alphabets, we
get a joint alphabet called hydropathy conformational letter
(hp-CL). A joint substitution matrix with $(17*2)^2$ indices is
derived from FSSP. Moreover, we check the three coding systems, say
AA, CL and hp-CL against a large database consisting proteins from
family to fold, with their performance on the TopK accuracy of both
similar fragment pair (SFP) and the neighbor of aligned fragment
pair (AFP). The TopK selection is according to the score calculated
by the coding system's substitution matrix. Finally, embedding hp-CL
in a pairwise alignment algorithm, say CLeFAPS, to replace the
original CL, will get an improvement on the HOMSTRAD benchmark.

{\bf Contact*:} wangsheng@itp.ac.cn
\end{footnotesize}
\end{abstract}
\maketitle















\section{Introduction}
Proteins fold into specific spatial conformations to perform their
biological functions \cite{ProteinStructure} and there are abundant
evidences to show their amino acid (AA) sequences determining the
structures \cite{Prediction1}. The attempt to find the relationship
between structure and sequence is a fundamental task in
computational biology \cite{SequenceStructure}.

Compared to the sequence world which is discrete of 20 AAs, the
structure world is continuous, though the local conformational space
of a protein backbone fragment is rather limited
\cite{LocalStructure}. The idea of representing the backbone with a
string of discrete letters was first observed by Corey and Pauling
\cite{PaulingAlpha,PaulingBeta} and later refined into the concept
of protein secondary structure elements (SSEs). However, segments of
a single SSE may vary significantly in their 3D structures,
especially for the state coil, which is not a true secondary
structure but is a class of conformations that indicate the absence
of regular SSEs, say alpha helix or beta strand \cite{ConfoLettApp}.
Although the SSE can be predicted with high accuracy ($\ge$80\%)
\cite{Prediction2}, the description of a protein in terms of its
SSEs is not sufficient to capture accurately its 3D geometry
\cite{Classify}.

To overcome this limitation, several groups have proposed the idea
that representing protein structures as a series of overlapping
fragments, each labeled with a symbol, which defines a structural
alphabet (SA) for proteins
\cite{WodakSA,BergSA,ProteinBlock,LevittSA,TufferySA,3D-BLAST}. Such
alphabet can be used to predict local structure
\cite{deBrevern1,deBrevern2,deBrevern3}, to reconstruct the
full-atom representation \cite{Reconstruct}, to identify the
structural motifs \cite{MetalBind}, to classify protein structures
\cite{Classify} and to search against a database
\cite{search1,search2}. We've proposed our SA, namely conformational
letter (CL) \cite{ConfoLett}, which is composed of 17 alphabets and
each with 4 residues in length. Our SA is focused on the fast
pairwise \cite{CLePAPS}, multiple \cite{CLeMAPS} and flexible
\cite{CLeFAPS} structure alignment problems, combined with its
substitution matrix CLESUM \cite{ConfoLett}.

After we discretized the continuous structure world into SAs, it is
the time to consider both AA and SA in a joint space. However, such
space is too large for about $(20*17)^2$ parameters when using the
current popular SAs. It is necessary to employ the reduction of AAs
\cite{ChanRed}. Several groups have put forward their reduced AAs
either experimentally or computationally. For example, Baker {\it
et.al} found a five-letter alphabet for 38 out of 40 selected sites
of SH3 chain \cite{BakerRed}; Wang \& Wang \cite{WWRed} introduced
the {\it minimal mismatch} principle to reduce the alphabet based on
Miyazawa-Jernigan's residue-residue statistical potential
\cite{MJmatrix}; Murphy {\it et.al} \cite{MWLRed} approached the
same problem using the BLOSUM matrix \cite{HHmatrix}. Recently, de
Brevern {\it et.al} proposed to use their SA, namely Protein Blocks
\cite{ProteinBlock} to analyze equivalences between the different
kinds of amino acids, then obtained their reduced AAs
\cite{deBrevernRed}.

Here we present a novel reduction method, called {\it entropic
clustering} \cite{ZhengRed}. Briefly, given two discrete
distributions A and B, merging $a_i$ and $a_j$ into one group
$a_{i\&j}$ will result in a loss of mutual information of A and B.
Thus, mutual information $I$ can be naturally chosen as the
objective function for optimized clustering. When grouping the 20
AAs into two categories, we've got a result of hydrophobic and
hydrophilic, which agrees with HP-model \cite{Hydropathy} exactly.
Then we construct a joint substitution matrix HP-CLESUM with
$(17*2)^2$ indices by the similar means as constructing CLESUM.

The following tests are employed to check and compare different
coding systems, namely AA, CL and hp-CL with their corresponding
substitution matrix, say BLOSUM, CLESUM and HP-CLESUM. We first
compare the TopK accuracy of SFPs (similar fragment pairs) and the
neighbor of AFP (aligned fragment pairs) against a large dataset
encompassing the protein homologous levels from family to fold
according to SCOP \cite{SCOP}; then we embed hp-CL into CLeFAPS
\cite{CLeFAPS}, replacing the original CL, to get an improvement
against the popular benchmark HOMSTRAD \cite{HOMSTRAD} .


\begin{figure*}[htb]
\centering
\includegraphics[width=0.8\textwidth]{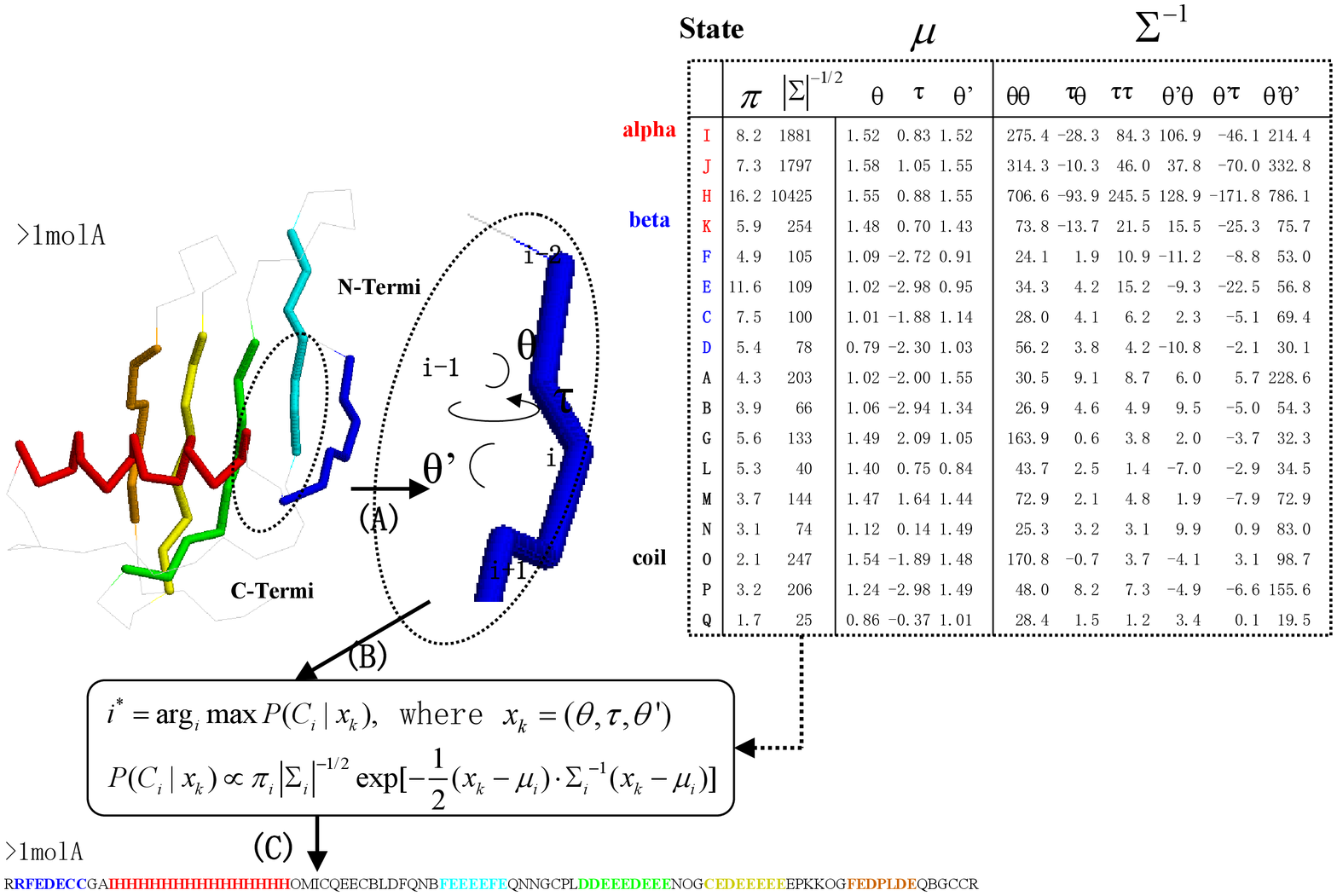}
\caption{{\scriptsize The conversion from 3D protein structure to 1D
CL string. (A) Given a sliding window of length four, may we get
four contiguous $C_\alpha$ atoms, and determine two bending angles
$\theta ,\theta '$ and a torsion angle $\tau $. (B) Select a state
which maximize the given data, where each state is a gaussian
distribution in the three dimension space of $\theta ,\tau ,\theta
'$. (C) Assign the state (letter) to the third place of the four
$C_\alpha$ atoms, so the $N$ length protein will get $N-3$ letters
finally, and we assign the first two and the last position with a
'blank' letter R.}} \label{figXYZtoCL}
\end{figure*}

\section{Materials and method}
\subsection{Datasets}
We use PDB-SELECT databank \cite{PDB-Select} to construct our CLs,
and use FSSP database \cite{FSSP} to derive the substitution matrix
CLESUM. Particularly, PDB-SELECT contains 1544 non-membrane proteins
from PDB \cite{PDB} with amino acid identity less than 25\%. FSSP is
based on exhaustive all-against-all structure comparison of the
representative protein structures, where the representative set
contains no pair which has more than 25\% sequence identity. A tree
for the fold classification of the 2,860 representative set is
constructed by a hierarchical clustering method based on the
structural similarities. Family indices of the FSSP are obtained by
cutting the tree at levels of 2, 4, 8, 16, 32 and 64 standard
deviations above the database average.


\begin{figure*}[htb]
\centering
\includegraphics[width=0.6\textwidth]{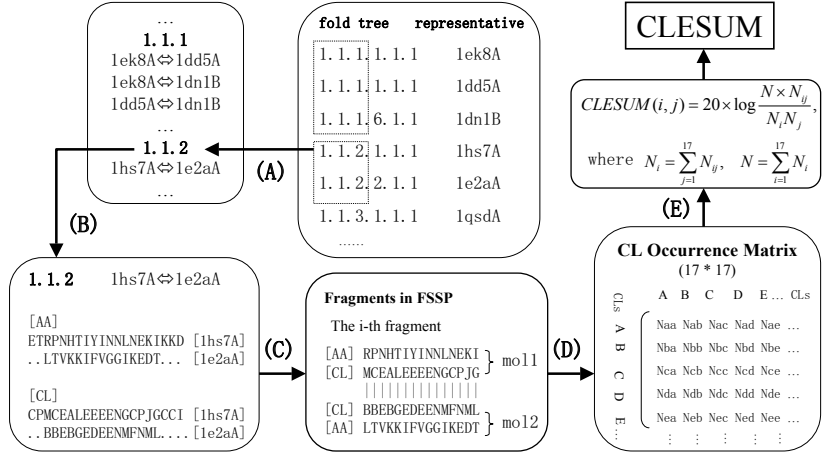}
\caption{{\scriptsize Flowchart of the construction of CLESUM. (A)
Collect all pairwise structures with the same first three family
indices (DALI Z-Score $\ge$ 8) in the representative set from FSSP.
(B) For each pair, extract the alignment. (C) For each alignment,
extract the ungapped aligned fragment. Each fragment contains both
AAs and CLs. (D) Count the occurrence number of CL duad. For
example, Nab means the total number of CL A and B occurred in the
alignment. (E) Calculate CLESUM from the occurrence matrix.}}
\label{figCLESUM}
\end{figure*}

\subsection{Conformational letter and its substitution matrix}
Four contiguous $C_\alpha$ atoms, say $i-2$, $i-1$, $i$ and $i+1$,
determine two bending angles $\theta ,\theta '$ and a torsion angle
$\tau $ which is the dihedral angle between the two planes of
triangles $i-2,i-1,i$ and $i-1,i,i+1$ (see Fig. \ref{figXYZtoCL}).
By using a mixture model for the density distribution of the three
angles, the local structural states from PDB-SELECT have been
clustered as 17 discrete states (see our previous work
\cite{ConfoLett} for more details). To use our SAs directly for the
structural comparison, a score matrix similar as BLOSUM
\cite{HHmatrix} for AAs is desired. In details, we first convert the
structures of the representative set from FSSP to their CL strings;
then collect all the pair alignments with the same first three
family indices (DALI Z-Score $\ge$ 8) (see Fig. \ref{figCLESUM});
finally count all ungapped aligned pairs of CLs to generate the
substitution matrix, say CLESUM (Table \ref{tableCLESUM}). The total
number of structures is 10,047 pairs, consisting of 175,723 fragment
pairs and 1,284,750 code pairs.

\begin{table}[htb]
\caption{CLESUM: The conformation letter substitution matrix (in
0.05 bit units).} \label{tableCLESUM} {\tiny\tt
\begin{tabular}{c rrrrrrrrrrrrrrrrr}\hline
$J$& 37&    &    &    &    &    &    &   &   &   &   &   &   &   &   &   &   \\
$H$& 13&  23&    &    &    &    &    &   &   &   &   &   &   &   &   &   &   \\
$I$& 16&  18&  23&    &    &    &    &   &   &   &   &   &   &   &   &   &   \\
$K$& 13&   5&  21&  49&    &    &    &   &   &   &   &   &   &   &   &   &   \\
$N$& -2& -34& -11&  28&  90&    &    &   &   &   &   &   &   &   &   &   &   \\
$Q$&-44& -87& -62& -24&  32&  90&    &   &   &   &   &   &   &   &   &   &   \\
$L$&-32& -62& -41&  -1&   8&  26&  74&   &   &   &   &   &   &   &   &   &   \\
$G$&-21& -51& -34& -13&  -8&   8&  29& 69&   &   &   &   &   &   &   &   &   \\
$M$& 16&  -4&   1&  12&   7&  -7&   5& 21& 61&   &   &   &   &   &   &   &   \\
$B$&-57& -96& -74& -50& -11&  12& -12& 13&-13& 51&   &   &   &   &   &   &   \\
$P$&-34& -60& -49& -36&  -3&   7& -12&  5&  8& 42& 66&   &   &   &   &   &   \\
$A$&-23& -45& -31& -19&  10&  16& -11& -6& -2& 20& 35& 73&   &   &   &   &   \\
$O$&-24& -55& -34&   5&  15& -13&  -4& -1&  5&-12&  4& 25&104&   &   &   &   \\
$C$&-43& -77& -56& -33&  -5&  29&   0& -4&-12&  7&  4& 13&  3& 53&   &   &   \\
$E$&-93&-127&-108& -84& -43&  -6& -21&-22&-47& 15& -5&-25&-48&  3& 36&   &   \\
$F$&-73&-107& -88& -69& -32&   3& -16& -5&-33&  7&  0&-20&-30& 20& 26& 50&   \\
$D$&-88&-124&-105& -81& -44&  14& -22&-31&-49& 13&-10&-17&-42& 21& 22& 21& 52\\
 & $J$ &  $H$ &   $I$ & $K$ & $N$ & $Q$ & $L$ & $G$ & $M$ & $B$ & $P$ & $A$ & $O$ &$C$  &$E$ &$F$ & $D$ \\
\hline\end{tabular}}
\end{table}


\begin{figure*}[htb]
\centering
\includegraphics[width=0.6\textwidth]{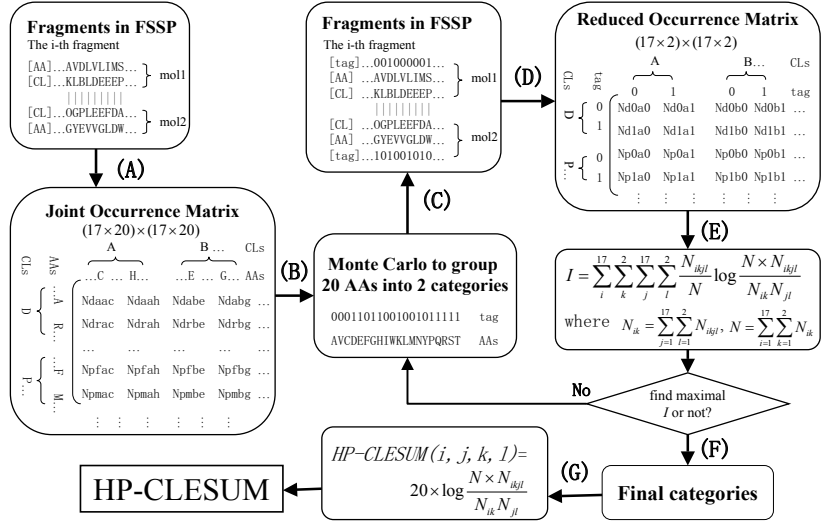}
\caption{{\scriptsize Flowchart of entropic clustering on the joint
space of both AAs and CLs. (A) For each fragment in FSSP, count the
pairwise number of joint occurrence. For example, Ndaac means the
total pairwise number of 'da' with 'ac', in each duad the former is
CL and the latter is AA. (B) Use Monte Carlo to randomly group 20
AAs into two categories. (C) Given an AA category, assign each AA to
its group. (D) Calculate the reduced occurrence matrix, here Nd0a1
means the total pairwise number of 'd0' with 'a1', in each duad the
former is CL and the latter is AA's tag. (E) Calculate the mutual
information of the reduced matrix. (F) If the categories which
maximize $I$ have been found, break the Monte Carlo recursion. (G)
Calculate the HP-CLESUM.}} \label{figHP-CLESUM}
\end{figure*}

\subsection{Entropic clustering}
From FSSP, which contains also the AA information, it is possible to
construct a substitution matrix in the joint space of the structure
and sequence. However, such matrix would have about
(17$\times$20)$\times$(17$\times$20) parameters (Fig.
\ref{figHP-CLESUM}(A)). If we group the 20 AAs into two clusters,
then the parameters of the matrix are reduced to
(17$\times$2)$\times$(17$\times$2).

Generally, the mutual information $I$ of two discrete distribution
($A$ and $B$) is defined as,
\begin{equation}
I = \sum\limits_{a,b}^{a \in A,b \in B} p(a,b)\log
\frac{p(a,b)}{p(a)p(b)}.
\end{equation}
If we cluster $a_i$ and $a_j$ into $a_{i\&j}$ leads to
\begin{eqnarray}
p(a_{i\&j} ) &=& p(a_i ) + p(a_j ),
\nonumber\\
p(a_{i\&j} ,b) &=& p(a_i ,b) + p(a_j ,b).
\end{eqnarray}
The difference between values of $I$ after and before clustering is
given by
\begin{eqnarray} \label{eqMutualMinus}
{[p(a_i ,b) + p(a_j ,b)]\log \frac{{p(a_i ,b) + p(a_j ,b)}}{{[p(a_i
) + p(a_j )]p(b)}}}
\nonumber\\
{ - p(a_i ,b)\log \frac{{p(a_i ,b)}}{{p(a_i )p(b)}} - p(a_j ,b)\log
\frac{{p(a_j ,b)}}{{p(a_j )p(b)}}}
\end{eqnarray}
which, by introducing
\begin{eqnarray}
x_i  &=& \frac{p(a_i ,b)}{p(a_i )},
\nonumber\\
\omega_i  &=& \frac{p(a_i )}{p(a_i ) + p(a_j )},
\end{eqnarray}
and their analogs $x_j$ and $\omega_j$, then defining $\left\langle
{F(x)} \right\rangle  = \omega_i F(x_i ) + \omega_j F(x_j )$ and
$F(\left\langle x \right\rangle ) = F(\omega_i x_i  + \omega_j x_j
)$ where $\omega_i  + \omega_j  = 1$. We may now see that Eq.
(\ref{eqMutualMinus}) is proportional to $f(\left\langle x
\right\rangle ) - \left\langle {f(x)} \right\rangle$ with $f(x) =
x\log x$. From the Jensen's inequality, for the convex function
$x\log x$ here we have $f(\left\langle x \right\rangle ) \le
\left\langle {f(x)} \right\rangle$, so $I$ never increases after any
step of clustering.

That is to say, merging any two members into one cluster will result
in a loss of mutual information. To make the loss of mutual
information as small as possible, $I$ should be maximized, so it can
be naturally chosen as the objective function during clustering. We
call this approach {\it entropic clustering} \cite{ZhengRed}. If we
partition $n$ objects into $m_1$ and $m_2$ classes, where $m_2
> m_1$, it is easy to prove that the maximal $I$ at $m_2$ is always greater
than the maximal $I$ at $m_1$ \cite{ConfoLettApp}.

Now turning back to our substitution matrix in the joint space, we
may define the average mutual information as follows like BLOSUM,
\begin{equation}\label{eqMutualInfo}
I = \sum\limits_{X,Y} {P_{XY} \log \frac{{P_{XY} }}{{P_X P_Y }}},
\end{equation}
where $X$ and $Y$ means a joint state of CL and AA (either reduced
or not). Given a clustering group may we calculate its $I$ based on
Eq. (\ref{eqMutualInfo}) and according to entropic clustering we
should get a categories which maximize $I$ (Fig.
\ref{figHP-CLESUM}(F)).


\section{Result}
\subsection{Joint substitution matrix of conformational letters
and reduced amino acids}

For clustering the 20 AAs into two categories, the Monte Carlo finds
{\tt AVCFIWLMY} and {\tt DEGHKNPQRST} as the groups, which is just
the hydrophobic and hydrophilic cluster \cite{Hydropathy}. Such
enlarged CLs are called hp-CLs.

\begin{table}[htb]
\caption{CLESUM-hh (lower left) and CLESUM-pp (upper right) (in
units of 0.05 bit): 'black dot' and 'white cycle' to indicate the
the CLs with different hydropathy AA types.} \label{tableHP-CLESUM1}
{\tiny\tt
\begin{tabular}{c rrrrrrrrrrrrrrrrr c}\hline
 & $\mathring{J}$& $\mathring{H}$& $\mathring{I}$& $\mathring{K}$& $\mathring{N}$& $\mathring{Q}$& $\mathring{L}$& $\mathring{G}$&
 $\mathring{M}$& $\mathring{B}$& $\mathring{P}$& $\mathring{A}$& $\mathring{O}$& $\mathring{C}$& $\mathring{E}$& $\mathring{F}$& $\mathring{D}$& \\
         &   45&   19&    24&    19&    3&  -41&  -27&  -13&    21&  -49&  -32&  -21&  -22&  -37&  -79&  -62&  -73&  $\mathring{J}$\\
         &     &   30&    24&     7&  -31&  -76&  -60&  -45&    -5&  -84&  -57&  -46&  -55&  -69& -113&  -92& -108&  $\mathring{H}$\\
$\dot{J}$&   40&     &    31&    27&   -6&  -57&  -38&  -28&     3&  -64&  -44&  -30&  -31&  -48&  -93&  -72&  -89&  $\mathring{I}$\\
$\dot{H}$&   18&   30&      &    57&   39&  -16&    5&   -8&    20&  -43&  -33&  -12&    6&  -24&  -66&  -52&  -64&  $\mathring{K}$\\
$\dot{I}$&   20&   26&    30&      &   94&   39&   15&   -5&    11&  -12&   -1&   14&   16&    0&  -35&  -25&  -31&  $\mathring{N}$\\
$\dot{K}$&   18&   14&    28&    59&     &   95&   33&   14&    -4&   10&    8&   21&  -14&   33&    3&   13&   20&  $\mathring{Q}$\\
$\dot{N}$&   -2&  -16&     3&    40&  108&     &   76&   35&    12&  -12&   -8&   -3&   -4&    7&  -11&   -5&  -14&  $\mathring{L}$\\
$\dot{Q}$&  -41& -110&   -67&   -19&   27&  109&     &   80&    33&    7&    9&   -1&   -4&    1&  -11&    5&  -17&  $\mathring{G}$\\
$\dot{L}$&  -33&  -58&   -37&    10&    8&   30&   86&     &    69&   -6&   13&    4&    9&   -4&  -32&  -21&  -34&  $\mathring{M}$\\
$\dot{G}$&  -22&  -48&   -35&   -10&  -11&    2&   40&   80&      &   64&   55&   31&   -4&   13&   20&   14&   25&  $\mathring{B}$\\
$\dot{M}$&   24&    7&     9&    16&   10&   -5&    9&   25&    71&     &   74&   38&    7&    9&    3&    7&    0&  $\mathring{P}$\\
$\dot{B}$&  -64&  -99&   -82&   -61&   -7&   15&   -3&   31&   -12&   59&     &   75&   31&   20&  -16&  -11&   -4&  $\mathring{A}$\\
$\dot{P}$&  -31&  -57&   -47&   -39&   -1&   13&  -21&    4&    13&   45&   76&     &  107&   10&  -37&  -24&  -32&  $\mathring{O}$\\
$\dot{A}$&  -19&  -36&   -26&   -25&   13&   11&  -24&  -10&    -6&   21&   42&   85&     &   60&   12&   28&   33&  $\mathring{C}$\\
$\dot{O}$&   -9&  -37&   -18&   -14&   25&   29&   -8&   13&    11&   13&   36&   32&  121&     &   45&   32&   30&  $\mathring{E}$\\
$\dot{C}$&  -49&  -87&   -67&   -45&  -17&   36&   -8&   -7&   -24&   10&    7&   17&   14&   62&     &   58&   30&  $\mathring{F}$\\
$\dot{E}$& -110& -138&  -126&   -98&  -56&   -5&  -24&  -22&   -58&   24&    3&  -23&  -22&    9&   44&     &   61&  $\mathring{D}$\\
$\dot{F}$&  -92& -131&  -105&  -106&  -46&    3&  -24&   -4&   -45&   10&    3&  -24&    5&   29&   34&   61&     &  \\
$\dot{D}$&  -98& -138&  -111&   -95&  -67&   33&  -20&  -36&   -66&   14&  -10&  -22&  -31&   26&   30&   28&   63&  \\
 & $\dot{J}$& $\dot{H}$& $\dot{I}$& $\dot{K}$& $\dot{N}$& $\dot{Q}$& $\dot{L}$& $\dot{G}$& $\dot{M}$&
 $\dot{B}$& $\dot{P}$& $\dot{A}$& $\dot{O}$& $\dot{C}$& $\dot{E}$& $\dot{F}$& $\dot{D}$&  \\
\hline\end{tabular}}
\end{table}

\begin{table}[htb]
\caption{CLESUM-hp (row-column) (in units of 0.05 bit).}
\label{tableHP-CLESUM2} {\tiny\tt
\begin{tabular}{c rrrrrrrrrrrrrrrrrr}\hline
$\mathring{J}$&  27&    3&    6&   3&  -3& -47& -38& -31&   6&  -65& -40& -28&  -8&  -50&  103&  -84&  101\\
$\mathring{H}$&  10&   14&   10&  -1& -25& -86& -69& -60& -11& -112& -67& -46& -35&  -87& -140& -126& -139\\
$\mathring{I}$&  13&    9&   13&  13&  -9& -64& -44& -40&  -7&  -80& -59& -30& -21&  -65& -119& -108& -121\\
$\mathring{K}$&  11&   -2&   12&  36&  23& -37& -15& -19&   4&  -48& -40& -25&   8&  -40&  -94&  -77&  -93\\
$\mathring{N}$& -12&  -48&  -25&   4&  76&  16& -10&  -8&  -1&   -9&  -6&   3&  15&  -13&  -46&  -41&  -59\\
$\mathring{Q}$& -50&  -94&  -66& -37&  14&  70&  11&   5& -13&   16&   5&  12&  -9&   15&  -14&  -11&   -5\\
$\mathring{L}$& -35&  -65&  -47&  -8&  12&  25&  65&  17&  -3&  -13&  -8& -10&  10&   -4&  -26&  -24&  -29\\
$\mathring{G}$& -20&  -54&  -35& -15& -17&   6&  16&  52&  12&   -5&   0&  -8&  21&   -5&  -31&  -13&  -37\\
$\mathring{M}$&  11&  -13&   -5&   4&  12&  -1&  -3&   9&  47&  -17&   4&  -4&  13&  -16&  -54&  -41&  -54\\
$\mathring{B}$& -54&  -95&  -75& -53& -19&  -1& -19&   5& -21&   34&  35&  18&   4&    6&   -2&   -3&   -1\\
$\mathring{P}$& -34&  -62&  -50& -36&  -6&   0& -19&   4&   1&   21&  53&  32&  17&   -1&  -22&  -16&  -27\\
$\mathring{A}$& -26&  -49&  -37& -24&   7&   6& -19&  -8&  -7&    2&  26&  66&  26&    2&  -37&  -30&  -31\\
$\mathring{O}$& -35&  -62&  -43&   4&   9& -21&  -7&  -4&  -3&  -35& -12&   0&  67&  -22&  -69&  -55&  -57\\
$\mathring{C}$& -42&  -74&  -57& -36& -10&  30&  -4&  -8& -12&   -3&   1&   7&   9&   42&  -14&    0&    2\\
$\mathring{E}$& -81& -116&  -95& -80& -44&  -6& -23& -24& -43&   12&  -1& -21& -19&    8&   25&   18&   14\\
$\mathring{F}$& -64&  -95&  -79& -62& -20&   2& -15&  -9& -30&    4&   5& -18&   2&   21&   14&   33&   14\\
$\mathring{D}$& -84& -114& -100& -79& -30&  20& -25& -32& -45&   11&  -3& -17& -29&   23&    9&    9&   34\\
 & $\dot{J}$& $\dot{H}$& $\dot{I}$& $\dot{K}$& $\dot{N}$& $\dot{Q}$& $\dot{L}$& $\dot{G}$& $\dot{M}$&
 $\dot{B}$& $\dot{P}$& $\dot{A}$& $\dot{O}$& $\dot{C}$& $\dot{E}$& $\dot{F}$& $\dot{D}$\\
\hline\end{tabular}}
\end{table}

The substitution matrix of hp-CLs is called HP-CLESUM, this symmetry
matrix can be divided into three sub-matrices: CLESUM-hh, CLESUM-pp,
and CLESUM-hp. The first two, shown in Table \ref{tableHP-CLESUM1},
correspond to the same hydropathy aligned amino acid types (i.e.,
{\it h-h} and {\it p-p}). The third, shown in Table
\ref{tableHP-CLESUM2}, corresponds to the different hydropathy types
{\it h-p}. As expected, compared with the original CLESUM (Table
\ref{tableCLESUM}), elements of CLESUM-hh and CLESUM-pp generally
become larger in absolute values, and those of CLESUM-hp show the
opposite tendency. The tendency is stronger for letters dominated by
helices or sheets.

\begin{table*}[htb]
\caption{TopK accuracy check with different strategies, from TopK
SFP check (left part) to TopK AFP's neighbor check (right part);
different coding systems, from AA (A\%), CL (C\%) to hp-CL (H\%);
different homologous level, from family (Fam), superfamily (Sup) to
fold; and different SFP length $L$, from 6, 12 to 18.}
\label{tableTopK-Test}{\scriptsize
\begin{tabular}{c|c| lccc lccc lccc lccl ||c| lccc lccc lccc}
\hline &&&\multicolumn{14}{c}{TopK SFP's Accuracy}&&&&\multicolumn{11}{c}{TopK AFP's Neighbor Accuracy}\\
\cline{4-17}\cline{21-31} Level&TopK&&
\multicolumn{3}{c}{$L=6$}&&\multicolumn{3}{c}{$L=12$}&&\multicolumn{3}{c}{$L=18$}&&
\multicolumn{2}{c}{Self(9-18)$^*$}&&TopK&&
\multicolumn{3}{c}{$L=6$}&&\multicolumn{3}{c}{$L=12$}&&\multicolumn{3}{c}{$L=18$}\\
\cline{4-6}\cline{8-10}\cline{12-14}\cline{16-17}
\cline{21-23}\cline{25-27}\cline{29-31}&&& {\tiny A\%}&{\tiny
C\%}&{\tiny H\%}&&{\tiny A\%}&{\tiny C\%}&{\tiny H\%}&& {\tiny
A\%}&{\tiny C\%}&{\tiny H\%}&&{\tiny C\%}&{\tiny H\%}&&&&
{\tiny A\%}&{\tiny C\%}&{\tiny H\%}&&{\tiny A\%}&{\tiny C\%}&{\tiny H\%}&&{\tiny A\%}&{\tiny C\%}&{\tiny H\%}\\
\hline\hline Fam
&1  && 53.1& 63.1& 68.0&& 62.4& 74.4& 77.6&& 62.4& 74.3& 77.1&& 73.1& 75.5&&1&& 25.8& 41.4& 46.3&& 46.1& 69.7& 74.2&& 62.4& 86.0& 88.5\\
&5  && 74.4& 86.7& 88.9&& 79.4& 90.7& 92.3&& 80.1& 91.2& 92.4&& 90.1& 91.6&&2&& 35.1& 53.3& 59.0&& 55.8& 79.3& 83.4&& 71.0& 91.7& 93.7\\
&10 && 82.0& 93.1& 94.4&& 85.7& 95.4& 96.2&& 86.6& 96.0& 96.6&& 95.2& 96.0&&3&& 41.1& 59.9& 65.9&& 61.4& 83.4& 87.3&& 75.4& 93.9& 95.6\\
&20 && 89.1& 97.1& 97.8&& 91.8& 98.2& 98.6&& 92.8& 98.4& 98.8&& 98.0& 98.5&&4&& 45.7& 64.4& 70.5&& 65.2& 85.9& 89.6&& 78.3& 95.0& 96.6\\
\hline Sup
&1  && 39.1& 48.7& 52.4&& 44.8& 58.3& 61.8&& 43.8& 58.2& 61.4&& 56.2& 59.4&&1&& 22.5& 36.3& 40.6&& 39.7& 63.9& 68.0&& 56.6& 83.3& 85.8\\
&5  && 58.2& 73.7& 76.7&& 62.4& 79.2& 81.8&& 63.0& 79.7& 81.5&& 78.4& 80.3&&2&& 31.0& 47.9& 52.9&& 48.9& 73.9& 77.6&& 65.1& 89.8& 91.7\\
&10 && 67.0& 83.1& 85.3&& 71.1& 87.3& 88.9&& 72.7& 87.9& 89.4&& 86.9& 88.5&&3&& 36.7& 54.7& 59.8&& 54.2& 78.5& 82.2&& 69.5& 92.3& 93.9\\
&20 && 77.2& 90.5& 91.8&& 81.0& 93.5& 94.5&& 82.8& 94.2& 95.0&& 93.5& 94.4&&4&& 41.1& 59.4& 64.7&& 57.9& 81.4& 84.9&& 72.5& 93.7& 95.2\\
\hline Fold
&1  && 13.9& 26.8& 29.7&& 17.0& 33.8& 36.7&& 17.0& 33.5& 35.6&& 31.9& 34.2&&1&& 10.7& 24.8& 27.0&& 20.0& 46.6& 49.8&& 32.2& 68.4& 70.7\\
&5  && 28.8& 50.3& 54.2&& 33.2& 57.0& 60.2&& 34.2& 58.6& 60.7&& 57.4& 59.6&&2&& 17.3& 35.1& 38.6&& 28.3& 57.8& 61.4&& 41.5& 77.4& 79.4\\
&10 && 39.5& 63.6& 66.7&& 45.7& 70.5& 73.0&& 48.4& 74.1& 75.4&& 72.5& 74.2&&3&& 22.3& 41.9& 46.0&& 33.7& 63.8& 67.6&& 46.9& 81.5& 83.5\\
&20 && 55.8& 77.2& 79.4&& 62.3& 84.5& 86.3&& 66.3& 87.0& 87.8&& 85.9& 87.0&&4&& 26.4& 46.9& 51.4&& 37.8& 67.8& 71.6&& 51.1& 84.3& 86.2\\
\hline \multicolumn{31}{l}{{\scriptsize $^*$: Self(9-18) means the
application of {\it self-adaptive} strategy \cite{CLeFAPS} of the SFP's length $L$ from 9 to 18. }}\\
\end{tabular}}
\end{table*}

\subsection{Comparison between different coding systems}
\subsubsection{Overview}

\begin{figure}[htb]
\centering
\includegraphics[width=1.0\columnwidth]{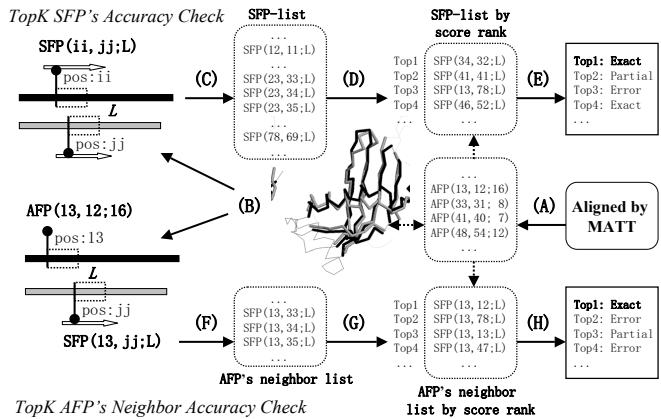}
\caption{{\scriptsize TopK accuracy check procedure. (A) Select a
pair of structures and aligned by MATT, resulting a series of AFPs.
(B) Encode the pairwise structures, the coding system may be AA or
SA. (C) Search for any SFPs with length $L$ according to a certain
coding system. (D) Sort these SFPs in descending order, the score is
calculated by the corresponding substitution matrix. (E) Check
whether there exists a correct SFP within TopK. (F) Search for any
possible neighbor of a given AFP whose length is over $L$ against
the other string. (G) Sort these AFP's neighbors. (H) Check their
accuracy.}} \label{figTopK-Test}
\end{figure}

A coding system in protein structure is defined as an alphabet
combined with its corresponding substitution matrix. Amino acids
(AA) or all kinds of SAs can be treated as coding systems, so long
as the alphabet has its substitution matrix. We'll compare the
performance of the following three ones, namely AA, CL and hp-CL,
based on their TopK accuracy against a benchmark (Fig.
\ref{figTopK-Test}). The difference between SFP and AFP is, we use
SFPs to describe all local similar fragment pairs, while AFPs is a
subset of SFPs that each of them should be in the final alignment
\cite{CLeFAPS}.

Note if the length of an AFP, say $Len$ is longer than $L$, we'll
check each positions and the total number is $Len-L+1$; if $Len$ is
shorter than $L$, we just skip this AFP. As a result, the TopK SFPs'
accuracy of a single pair of structures is a 0 or 1 measure, that is
to say, within TopK we find a correct SFP or not. While the TopK
neighbor of AFP's accuracy is calculated by summing all correct
positions found in any AFPs then dividing the total valid positions,
the result is between 0.0 to 1.0.

The benchmark we use is divided in three levels: family, superfamily
and fold according to SCOP \cite{SCOP}. In family set, we use all
SCOP families which have 2 to 25 members in ASTRAL 40\% compendium
\cite{ASTRAL} and the total pair number is 21,039. In superfamily
and fold, it is convenient to use SABmark \cite{SABmark} instead of
using ASTRAL 40\% because SABmark is systematically arranged and
elaborately checked at both superfamily and fold level. The
superfamily set contains 3,645 domains sorted into 426 subsets and
the fold set (or be called {\it twilight zone} \cite{TwilightZone})
contains 1,740 domains sorted into 209 subsets, where each subsets
contain between 3 to 25 structures. The superfamily set contains
18,724 structure pairs and the fold set contains 10,306. We apply
MATT \cite{MATT} to conduct all-against-all pairwise alignment
within each family or subset as our gold standard.

\subsubsection{Performance}
Table \ref{tableTopK-Test} shows that, hp-CL performs best while CL
follows the second, both of them outperforms 10\% to 100\% than AA.
For details, with the increase of TopK and length $L$, the accuracy
of all coding systems grows better, while from family to fold level,
the accuracy declines. It is surprising that at fold level, the
accuracy of hp-CL outperforms AA more than 50\% at the TopK SFP
accuracy test and more than 100\% at the TopK neighbor of AFP test,
while in the latter, hp-CL got the average accuracy at about 71\%
given $L=18$ from the Top-1 highest neighbor of an AFP. Such feature
may be applied to construct the Highest Similarity Fragment Block
(HSFB) during the multiple structure alignment \cite{ConfoLettApp}.
Given a seed structure and a certain position, if this position got
many high score neighbors in other structures, may we say that this
block (consisting the seed position and its neighbors) has a more
probable chance in the final multiple alignment.

\begin{figure}[htb]
\centering
\includegraphics[width=0.9\columnwidth]{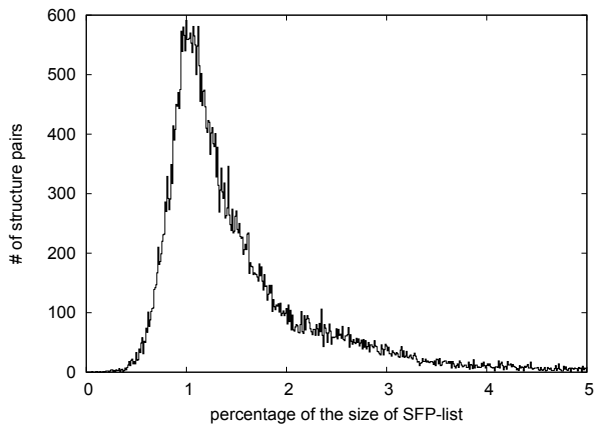}
\caption{{\scriptsize Histogram of the percentage of SFP-list's size
to the search space (moln1*moln2) between all homologous levels
(from family to fold) under the {\it self-adaptive} strategy, where
the moln1 (moln2) is the first (second) structure's size. The peak
value is about 1.045\% and the mean value is about 1.5505\%. The
total pair count is 50,069.}} \label{figHistogram}
\end{figure}

Moreover, we've shown the effectiveness of {\it parameter
self-adaptive} strategy to create the SFP-list in \cite{CLeFAPS}. At
most cases {\it self-adaptive} strategy is compatible with {\it
fixed parameters}, while the size of the SFP-list can be controlled
empirically to about O(n$^2$/LEN\_{}H/6) with the LEN\_{}H=9 (Fig.
\ref{figHistogram}). However, its hard to control the balance
between the size of the SFP-list and the threshold of SFP generated
with fixed parameters strategy. Actually, the data of fixed length
used in Table \ref{tableTopK-Test} is considered all O(n$^2$) SFPs,
then to select TopK; we've tested different SFP thresholds, if it is
set too high there'll lead to blank or few SFP-list while if it is
set too loose then the SFP-list will be too much (data not shown).

Finally, we may get the conclusion that, during structure
comparison, the only consideration of the TopK highest SFPs to built
the initial alignment is feasible from family level to fold, so long
as the coding system is specific enough. Also the employment of
parameter self-adaptive strategy to generate SFPs is effective and
economic.

\subsection{Implement of hydropathy conformational letters
to structural alignment}

We embed hp-CL to the pairwise protein structural alignment problem
under the framework of CLeFAPS \cite{CLeFAPS}. Particularly, we
first transform each structures to its hp-CL strings; then search
for both highly specific SFPs (SFP\_{}H) that have a high HP-CLESUM
score to build an initial alignment from the best TM-score
\cite{TMscore} SFP within TopK, and highly sensitive SFPs (SFP\_{}L)
that have a low HP-CLESUM score (must above 0) to refine the
alignment through fuzzy-add strategy. These two SFP-lists can be
generated simultaneously \cite{CLePAPS}; finally we apply an
elongation based on Vect-score to collect local flexible fragments.

\begin{table}[htb]
\caption{Alignment accuracy metric on HOMSTRAD}
\label{tableHOMSTRAD}
\begin{tabular}{c  c  c  c}
\hline
Metric & CLeFAPS(CL) \ \ & CLeFAPS(hp-CL) \ \ & MATT \\
\hline
C/LOA$^1$   &   0.929   &   0.939    &  0.948\\
C/LOR$^2$   &   0.898   &   0.907    &  0.831\\
\hline \multicolumn{3}{l}{{\footnotesize $^1$: Correct/(Length
of the algorithm).}}\\
\multicolumn{3}{l}{{\footnotesize $^2$: Correct/(Length
of the reference).}}\\
\end{tabular}
\end{table}

HOMSTRAD is a database of protein structural alignments for
homologous families \cite{HOMSTRAD}. Its alignments were generated
using structural alignment programs, then followed by a manual
scrutiny of individual cases. There are totally 1033 families (633
at pairwise level). We'll show the improvement based on hp-CL as the
coding systems instead of CL under the same algorithm, say CLeFAPS,
in Table \ref{tableHOMSTRAD}.

\section{Discussion and future work}
To explore the joint space of both AAs and SAs, entropic clustering
is a simple but effective approach. In this work, only the reduction
of AAs is considered, while we may also reduce CLs and AAs
simultaneously while balancing the accuracy and the parameter
numbers. For example, if reducing the CL to 9 letters, (actually,
from Fig. \ref{figXYZtoCL}, there are 4 codes for helix which can be
grouped to one cluster, the same as sheet.) we may then consider up
to four AA cluster instead of two, while the total alphabet number
is about the same as hp-CL.

It is interesting that, hp-CL can be applied during the situation we
know only a little information about the AA sequence of the
structure, i.e., the hydropathy features, or even none. That is true
because, from the knowledge of protein design
\cite{ProteinDesign2,ProteinDesign1}, the hydropathy patterns from a
3D structure may probably be deduced. Then the usage of hp-CLs and
HP-CLESUM that consider the hydropathy patterns will get a more
accurate result than CLs and CLESUM that only consider the 3D
structure.

We also verify a basic idea in CLeFAPS, i.e., {\it self-adaptive}
strategy to generate SFPs, that we needn't consider the parameters
to deal with different purposes and different proteins. The result
showed its accuracy is maintained well and the SFP-list size is
controlled in O(n$^2$/LEN\_{}H/6) while its hard to judge the
balance between accuracy and size with fixed parameters.

TopK accuracy check has demonstrated the basic strategy of both
CLePAPS and CLeFAPS efficient, which only considers TopK highest
SFPs to built the initial alignment. Moreover, TopK accuracy check
is an effective approach to measure the coding systems against a
reference dataset, especially to judge the substitution matrix. If a
coding system is good enough, it should rank those SFPs with highly
specificities top enough among other SFPs. In a future work, we'll
use this approach to test the current available SAs based on their
performance for finding specific SFPs. Also we can do the comparison
between SAs and RMSD values or some p-values derived from RMSD. Such
comparison between 1D coding systems with 3D expression will show
the effectiveness of SAs because they contain the statistic
information from the database \cite{CLePAPS}.

\section*{Acknowledgments}
This work is supported by ...

\end{document}